\documentclass[aps,pra,superscriptaddress,amsmath,amssymb,twocolumn]{revtex4}
\usepackage{amsmath}
\usepackage{amsfonts}
\usepackage{amsthm}
\usepackage{graphicx}
\usepackage{dcolumn}
\usepackage{stackrel,amssymb}
\usepackage{color}
\usepackage{soul}
\usepackage{ulem}
\usepackage{bm}

\newcommand{\ket}[1]{\vert#1\rangle}

\newcommand{\prt}[1]{\left(#1\right)}
\newcommand{\prtq}[1]{\left[#1\right]}
\newcommand{\prtg}[1]{\left\{#1\right\} }

\usepackage{graphicx}
\usepackage{hyperref}

\usepackage{multirow,array}

\usepackage[utf8]{inputenc}

\begin{document}

\title{ Microscopic and phenomenological models of driven systems in structured reservoirs }

\author{Gian Luca Giorgi}
\affiliation{Institut UTINAM - UMR 6213, CNRS, Universit\'{e} Bourgogne Franche-Comt\'{e}, Observatoire des Sciences de l'Univers THETA, 41 bis avenue de l'Observatoire, F-25010 Besan\c{c}on, France}

\affiliation{IFISC (UIB-CSIC), Instituto de F\'isica Interdisciplinar y
Sistemas
Complejos, UIB Campus, 07122, Palma de Mallorca, Spain}

\author{Astghik Saharyan}
\affiliation{Laboratoire Interdisciplinaire Carnot de Bourgogne, CNRS UMR 6303, Universit\'{e} Bourgogne Franche-Comt\'{e}, BP 47870, F-21078 Dijon, France}

\author{St\'ephane Gu\'erin}
\affiliation{Laboratoire Interdisciplinaire Carnot de Bourgogne, CNRS UMR 6303, Universit\'{e} Bourgogne Franche-Comt\'{e}, BP 47870, F-21078 Dijon, France}

\author{Dominique Sugny}
\affiliation{Laboratoire Interdisciplinaire Carnot de Bourgogne, CNRS UMR 6303, Universit\'{e} Bourgogne Franche-Comt\'{e}, BP 47870, F-21078 Dijon, France}

\author{Bruno Bellomo}
\email{bruno.bellomo@univ-fcomte.fr}
\affiliation{Institut UTINAM - UMR 6213, CNRS, Universit\'{e} Bourgogne Franche-Comt\'{e}, Observatoire des Sciences de l'Univers THETA, 41 bis avenue de l'Observatoire, F-25010 Besan\c{c}on, France}

\begin{abstract}
 We study the paradigmatic model of a qubit interacting with a structured environment and driven by an external field by means of a microscopic and a phenomenological model. The validity of the so-called fixed-dissipator (FD) assumption, where the dissipation is taken as the one of the undriven qubit is discussed. In the limit of a flat spectrum, the FD model and the microscopic one remarkably practically coincide.  For a structured reservoir, we show in the secular limit that steady states can be different from those determined from the FD model, opening the possibility for exploiting reservoir engineering. We explore it as a function of the control field parameters, of the characteristics of the spectral density and of the environment temperature. The observed widening of
the family of target states by reservoir engineering suggests new possibilities in quantum control protocols.

\end{abstract}

\maketitle

\section{Introduction}

Generally speaking, no quantum system can be considered as completely isolated from its environment, which is at the origin of dissipation and decoherence~\cite{breuer,gardiner}. These dissipative processes could negatively influence control protocols which aim at bringing a quantum system towards a desired target state, such as the ones considered in quantum control~\cite{dalessandro,carlini,sugny,lapertprl,shapiro,glaser,koch} and in  remote state preparation~\cite{rsp,rsp2}.
Dissipative dynamics can be strongly modified by using, for instance, dynamical decoupling strategies~\cite{viola1,viola2,addis} or tuned into a useful tool, e.g. by properly engineering the characteristics of the environment, to generate specific states~\cite{Briegel2006, Polzik2011, Bellomo2013, Bellomo2015}.

The study of open quantum systems usually involves approximations~\cite{breuer}.
Master equations are often derived in the weak-coupling regime between the system and the bath (Born approximation) and for memory-less dynamics with time-independent dissipation rates (Markovian approximation). Other common assumptions concern the absence of initial correlation between the system and its environment and the secular approximation. A standard way to obtain a master equation is based on a microscopic approach which takes into account the full Hamiltonian of the system and the environment, including their mutual coupling, and by performing (some of) the approximations described above. The system dynamics are completely positive as long as the master equation is in the Lindblad form~\cite{lindblad,Gorini}.

An alternative route to take into account environmental effects relies on a phenomenological description of standard dissipative and dephasing mechanisms where Lindblad superoperators are designed to reproduce the desired process and the dissipator is built ``by hands", for instance by inferring the decay rates from experimental data. This approach may lead to a drastic simplification of the dynamics~\cite{breuer}.
However, there are scenarios for which the phenomenological technique must be taken with a pinch of salt, as it may not be able to capture all the relevant aspects of the dynamics. In particular, this problem becomes crucial when an  external control field is exerted to the system, or when different systems are coupled to each other while dissipating locally~\cite{adesso,naseem,marco}.

In the context of phenomenological modelling of open quantum systems subject to external control fields,  a standard assumption is the so-called fixed-dissipator (FD) assumption~\cite{Lacour,sauer,lapert,mukherjee,sauer2}. This is based on the hypothesis that the dissipative part of the master equation is not changed by the control term. Comparisons between phenomenological and microscopic master equations have been realized, also considering, in the case of bipartite systems, the effects due to a strong coupling between the internal parts~\cite{Scala2007,Rivas2010,Blais2011,Werlang2014,Bylicka2018}. Problematic consequences of phenomenologically derived master equations in quantum thermodynamics have been recently discussed~\cite{Levy2014, DeChiara2018, naseem}
and, as shown in Ref.~\cite{sabrina}, the FD assumption can be at the origin of non physical trajectories in the non-Markovian limit. A generalized approach trying to unify phenomenological and microscopic approaches has been recently proposed \cite{Shavit}.

The scope of this paper is to study the case of a driven qubit interacting with a structured environment by means of a microscopic model and to analyze the consequences of the FD assumption. This includes the possibility of using reservoir engineering as a tool for quantum control. For that purpose we mainly study the dynamics on asymptotic time scales and compare the steady states reachable with a microscopic master equation (MME) with the ones given by a master equation based on a fixed dissipator (FDME). We show that manipulating the environment through reservoir engineering, which is possible when the environment spectrum is not flat, allows one to obtain a collection of stationary states that can be very different from the ones given by the FDME.

The paper is organized as follows. In Sec.~\ref{sec2}, we introduce the model of a qubit driven by a monochromatic laser and interacting with a bosonic environment. In Sec.~\ref{sec3}, the MME for such a system is explicitly derived. Some technical details are reported in the Appendix\ref{appe}. In Sec.~\ref{secfd}, we review the FD assumption, while in Sec.~\ref{sec4}, we present the main results of this study in the case of structured environments, both at zero and non-zero temperatures. Discussion and prospective views are presented in Sec.~\ref{sec5}.

\section{The model system} \label{sec2}

For the sake of simplicity, we tackle the problem of comparing microscopic and phenomenological models of  driven systems in structured environments
by revisiting a simple quantum system made of a qubit of frequency $\omega_0$ (whose free Hamiltonian is aligned along the $z$-axis) driven by a monochromatic control laser field whose frequency is $\omega_L$  and whose initial  phase is $\varphi$~\cite{haikka, tanas}. We define the detuning as $\Delta=\omega_0-\omega_L$ and we refer to the Rabi frequency $\Omega$, related to the intensity of the laser field, as the driving amplitude. We assume henceforth that $\omega_0$ and $\omega_L$ are much larger than $\Delta$  and $\Omega$. The starting Hamiltonian is given by:
\begin{equation}
\bar{H}_S=\frac{\hbar\omega_0}{2}\sigma_z+\hbar\Omega\cos(\omega_L t+\varphi)\sigma_x,
\end{equation}
where $\sigma_z$ and $\sigma_x$ are Pauli matrices. Under the above condition on the parameters we may apply the rotating wave approximation on $\bar{H}_S$.  We also move to a frame rotating at frequency $\omega_L$, by means of the unitary operator $U_L=\exp\left[ -i (\omega_L t+\varphi) \sigma_z  /2 \right]$ (also absorbing the time-independent phase factor $\varphi$), obtaining:
\begin{equation}\label{eqh}
H_S=\frac{\hbar\Delta}{2}\sigma_z+\frac{\hbar\Omega}{2}\sigma_x.
\end{equation}
The interaction between the system and the environment, which is assumed not to depend on the control field (see, for instance, Ref.~\cite{weiss}), reads as follows:
\begin{equation}
\bar{H}_I=\sum_k  \hbar\left( g_k a_k+ g_k^* a_k^\dag\right)\sigma_x,
\end{equation}
where $a_k$ $\prt{a_k^\dag}$ are the annihilation (creation) operators of the bosonic bath and $g_k$ are the coupling constants.
In the above rotating frame, $\bar{H}_I$ is transformed into:
\begin{equation}\label{hi}
H_I=\sum_k  \hbar \prt{g_k a_k+ g_k^* a_k^\dag}\left[ e^{i (\omega_L t+\varphi)}\sigma_+ +e^{-i (\omega_L t+\varphi)}\sigma_-\right].
\end{equation}
The free Hamiltonian of the environment has the form $H_E=\sum_k \hbar \omega_k a_k^\dag a_k$.
$H_S$ can be diagonalized  as $H_S=\frac{\hbar \nu}{2} (|\phi_+\rangle\langle \phi_+| -|\phi_-\rangle\langle \phi_-|)$,
with $\nu=\sqrt{\Delta^2+\Omega^2}$. Its eigenstates are
\begin{eqnarray}
 |\phi_+\rangle=C |e\rangle+S  |g\rangle,\nonumber\\
 |\phi_-\rangle=C |g\rangle- S  |e\rangle, \label{fi}
\end{eqnarray}
where $|g\rangle$ and $|e\rangle$ are, respectively, the ground and the excited state of the quibt free Hamiltonian $(\hbar \omega_0/2)\sigma_z$, $C=\cos(\theta/2) $, $S=\sin(\theta/2)$ and
\begin{equation}\label{theta}
\theta=2 \arctan[(\nu-\Delta)/\Omega ].
\end{equation}
For example, for a given $\Delta>0$, $\theta$ goes from 0 to $\pi/2$ when $\Omega$ goes from 0 to infinity.

\section{Microscopic master equation}\label{sec3}
To derive a microscopic master equation, the qubit driven by the field is treated first. The resulting dressed qubit is next coupled to the environment by
expressing $H_I$ in terms of the eigenoperators of $H_S$ and  the standard Born and Markov approximations are applied
 (see also Refs.~\cite{tanas,haikka}).

Defining $\tilde{\sigma}_z=|\phi_+\rangle\langle \phi_+|-|\phi_-\rangle\langle \phi_-|$ and $\tilde{\sigma}_\pm=|\phi_\pm\rangle\langle \phi_\mp|$,
we can express the operators entering $H_I$ in terms of eigenoperators of $H_S$ as:
\begin{eqnarray}\label{sx}
\sigma_\pm&=&C^2 \tilde{\sigma}_\pm-S^2\tilde{\sigma}_\mp+SC\tilde{\sigma}_z,
\nonumber\\
\sigma_z&=&\cos\theta \tilde{\sigma}_z-\sin\theta\tilde{\sigma}_x.
\end{eqnarray}
The detailed derivation of the MME is presented in the Appendix\ref{appe}. Its final form in the Schr\"odinger picture is:
\begin{equation}
\label{metot}
\dot \rho=- \frac{i}{\hbar}[H_S+ H_{LS},\rho]+{\cal D}^{\rm sec}(\rho)+{\cal D}^{\rm nsec}(\rho),
\end{equation}
where $H_{LS}$ is the Lamb shift Hamiltonian, whose role is discussed in the Appendix\ref{appe}, while ${\cal D}^{\rm sec}(\rho)$ and ${\cal D}^{\rm nsec}(\rho)$ are, respectively,  the secular  and the non-secular parts of the dissipator, the latter featuring terms oscillating at frequencies $\nu$ and $2\nu$.

With regards to ${\cal D}^{\rm sec}(\rho)$, it is given by:
\begin{equation}
{\cal D}^{\rm sec}(\rho)=\gamma_-^{\theta} {\cal L}\prtq{\tilde{\sigma}_+}(\rho)+\gamma_+^{\theta} {\cal L}\prtq{\tilde{\sigma}_-}(\rho)+\gamma_z^{\theta} {\cal L}\prtq{\tilde{\sigma}_z}(\rho),
\end{equation}
where the Lindblad superoperator is $ {\cal L}\prtq{\hat X}(\rho)=\hat X \rho \hat X ^\dagger -\prtg{\rho ,\hat X^\dagger \hat X }/2$,  with
\begin{eqnarray}\label{gammas}
\gamma_{-}^{\theta}&=&2 \pi \left\{ C^4 J(\omega_L+\nu) n(\omega_L+\nu) + S^4\ J(\omega_L-\nu)  \nonumber \right.  \\      &    \times  &   \left. [1+n(\omega_L-\nu)]  \right\},\nonumber\\
\gamma_{+}^{\theta}&=& 2 \pi \left\{ C^4 J(\omega_L+\nu)[1+n(\omega_L+\nu)]+S ^4 J(\omega_L-\nu) \nonumber  \right. \\ &\times&n(\omega_L- \nu)\left.\right\} \nonumber, \\
\gamma_{z}^{\theta}&=&  2 \pi  \left\{ S^2 C^2J(\omega_L)[1+2 n(\omega_L)]\right\},
\end{eqnarray}
where $J(\omega)$ is the spectral density of the environment, $n(\omega)=1/\prtq{\mathrm{e}^{\hbar\omega/(k_B T)}-1}$  is the average number of photons in the bath at frequency $\omega$, $k_B$ being the Boltzmann constant. The above coefficients can be rewritten as
\begin{eqnarray}\label{gammas2}
\gamma_{-}^{\theta}&=& C^4 \gamma_+ n_+ + S^4 \gamma_-(1+n_-),\nonumber\\
\gamma_{+}^{\theta}&=&  C^4 \gamma_+(1+n_+)+S ^4 \gamma_- n_- \nonumber, \\
\gamma_{z}^{\theta}&=&   S^2 C^2\gamma_0(1+2 n_0)],
\end{eqnarray}
where $\gamma_{p}=  2 \pi J(\omega_L+ p\nu)$ and $n_{p}= n(\omega_L+ p \nu)$, being $p=\{+ 1 , - 1, 0\}$ (for any parameter $l$ depending on $p$ we use the shorthand notation $l_p=\{l_+, l_-, l_0\}$).

The operator  ${\cal D}^{\rm nsec}(\rho) $ and its coefficients are reported in the Appendix\ref{appe}.
In Sec.~\ref{SecT0} we give some comments about when their effect can not be neglected. A detailed analysis of the limits of validity of the secular approximation in our system can be found in Ref.~\cite{Shen}.

\section{The reference case: the fixed dissipator}\label{secfd}
In Sec.~\ref{sec3}, we have seen that, in the microscopic approach, the dissipator depends on the control field acting on the qubit. The FD approach consists in neglecting this dependance and in assuming that the dissipative part of the master equation is equal to the one in the absence of the control field, i.e. the qubit coupled to the environment is treated first, and this single entity is next coupled to the laser. The application of this procedure is well-known in quantum control protocols (see e.g. Refs.~\cite{Lacour,sauer,lapert,mukherjee,sauer2}). Recently, this approach has been used to determine the control Hamiltonian that counteracts a given dissipation~\cite{sauer,sauer2}. In this context, we consider a density matrix evolving according to a general (Lindblad) master equation:
\begin{equation}\label{MEFD}
\dot{\rho}=-\frac{i}{\hbar}[H,\rho]+{\cal D}^{\rm fd}(\rho).
\end{equation}
The set of stationary solutions $\dot{\rho}^{\rm fd}=0$, which are compatible with the fixed dissipator ${\cal D }^{\rm fd}(\rho^{\rm fd})$ can be computed by disregarding the coherent part.
Since the coherent part of the master equation cannot change the spectrum and then the purity of the state,  the same must also be true for the dissipator~\cite{sauer,lapert,sauer2}. Then, the collection of stationary states $\rho^{\rm fd}$ must obey the relation:
\begin{equation}\label{fdeq}
\forall   n  \in \{2,\dots,d\}: \; \;{\rm Tr}\prtg{\prt{\rho^{\rm fd}}^{n-1}{\cal D }^{\rm fd}\prt{\rho^{\rm fd}}}=0,
\end{equation}
where $d$ is the dimension of the Hilbert space.
Thus, we have defined a (fixed) dissipator and a family of Hamiltonians. For any steady state, we can find the Hamiltonian $H$ such that
$\dot{\rho}^{\rm fd}=0$. Writing  the steady state as $\rho^{\rm fd}=\sum_{\alpha=1}^d \lambda_\alpha |\alpha\rangle\langle \alpha|$, it follows that~\cite{sauer,sauer2}:
\begin{equation}
H=\sum_{\alpha,\beta:\lambda_\alpha\neq \lambda_\beta}\frac{i \langle\alpha| D^{\rm fd}\prt{\rho^{\rm fd}}|\beta\rangle}{\lambda_\alpha-\lambda_\beta}|\alpha\rangle\langle\beta|.
\end{equation}
\begin{figure}[t!]
 % \centering
  \includegraphics[width=0.35 \textwidth]{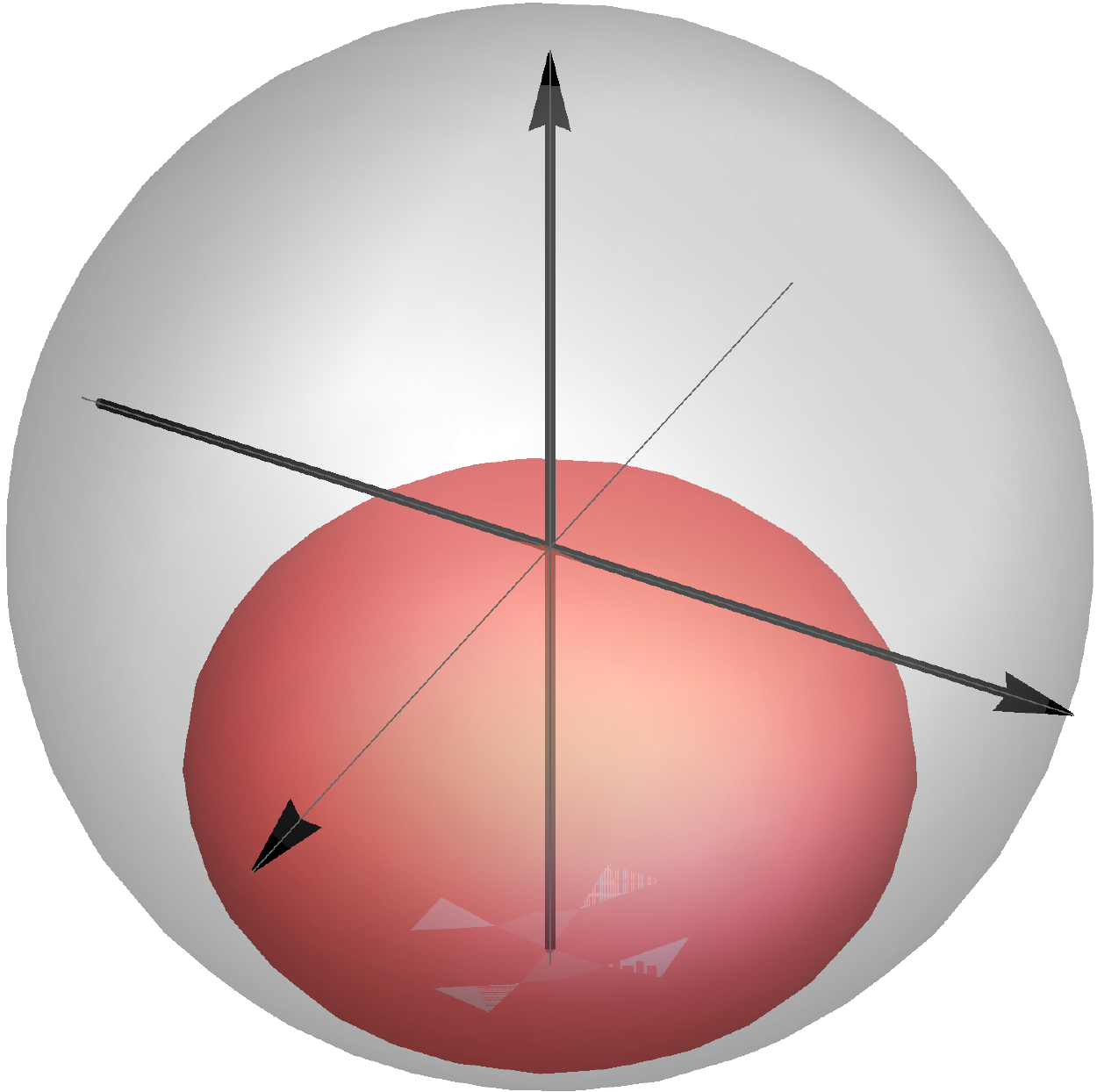}\\
  \caption{Bloch sphere (gray) and steady-state ellipsoid (red). The ellipsoid has been drawn by taking $T=0$.}\label{Figsp}
\end{figure}
For the case of the qubit introduced in Sec.~\ref{sec2}, we only need to satisfy ${\rm Tr}\prtg{\rho^{\rm fd}{\cal D }^{\rm fd}\prt{\rho^{\rm fd}}}=0$. In this model, at a given bath temperature $T$, the FD is equal to the dissipator one would obtain in the absence of the control field ($\Omega\rightarrow 0$). This can be obtained from the microscopic dissipator of Eq.~(\ref{metot}) taking $\theta=0$ and $\nu=\Delta$. In this limit, the decay rates of Eq.~\eqref{gammas2} tend to: $\gamma_{-}^{0}=\gamma_{\rm fd} n_{\rm fd}$, $\gamma_{+}^{0}=\gamma_{\rm fd}(1+n_{\rm fd})$, and $\gamma_{z}^0=0$, where $\gamma_{\rm fd}=2 \pi J(\omega_0)$ and $n_{\rm fd}= n(\omega_0)$. The fixed dissipator is then of the form:
\begin{equation}\label{FD}
D^{\rm fd}(\rho)=\gamma_{\rm fd} n_{\rm fd}  {\cal L}[\sigma_+](\rho)+\gamma_{\rm fd} (1+n_{\rm fd})  {\cal L}[\sigma_-](\rho).
\end{equation}
It follows that in the FD approach the steady state depends on $\gamma_{\rm fd}$ and it can be expressed, using $H=H_S$ in Eq.~\eqref{MEFD}, as (restoring the dependence on $\varphi$):
\begin{eqnarray}\label{ssfd}
\rho_{ee}^{\rm fd}&=& \frac{n_{\rm fd}}{1+2n_{\rm fd}}+   \frac{\Omega^2/(1+2 n_{\rm fd})}{\gamma_{\rm fd}^2(1+2 n_{\rm fd})^2+4 \Delta ^2+2 \Omega^2},\nonumber\\
\rho_{eg}^{\rm fd}&=& -\Omega \frac{2 \Delta/(1+2 n_{\rm fd})+i \gamma_{\rm fd}  }{\gamma_{\rm fd}^2(1+2 n_{\rm fd})^2+4 \Delta ^2+2 \Omega^2}e^{-i \varphi}. \label{regf}
\end{eqnarray}
The FD steady solutions by varying the control field parameters, $\Omega$, $\Delta$ and $\varphi$, are represented in Fig.~\ref{Figsp} (for $T=0$) where they are shown to lie on the surface of an ellipsoid inside the Bloch sphere~\cite{recht,sauer,lapert,sauer2}. This ellipsoid is a standard geometric structure in nuclear magnetic resonance~\cite{lapert,levitt,ernst}. For $T\neq 0$, the steady states lie on a smaller ellipsoid inside the one depicted in Fig.~\ref{Figsp}.

\section{Reservoir engineering through microscopic master equation with structured environment}\label{sec4}

We present in this section the control of steady states by using the MME of Sec.~\eqref{sec3}. In particular, we compare the steady state solutions of the FDME with the ones provided by the MME to discuss how the control of the system is modified when the environment is used as a tool to suitably tailor the asymptotic states. We also compare some specific dynamics to highlight our results.

In the case of flat spectrum, it holds $\gamma_+ =  \gamma_-= \gamma_0= \gamma_{\rm fd}$ and one can show the remarkable property that, under the approximation $n_+ \approx   n_- \approx n_0  \approx n_{\rm fd}$, the MME coincides exactly with the FDME~\cite{tanas} (see the Appendix\ref{appe} for a complete derivation): the steady states of the MME and of the FDME are thus  the same for any $T$. In particular, in the secular limit, in the $z$-basis of the frame rotating at the laser frequency (after restoring the phase $\varphi$), the MME steady solutions are equal to the ones of Eq.~\eqref{ssfd} after discarding the terms containing $\gamma_{\rm fd}$, which are indeed negligible in this limit. One can show that the geometric form of the steady state solutions obtained by varying the control field parameters, $\Omega$, $\Delta$ and $\varphi$, corresponds to the very same ellipsoid of Fig.~\ref{Figsp}. When non-secular terms are added, the microscopic steady states coincide with the ones obtained with the FD, given in Eq.~\eqref{ssfd}. We consider below the case of structured environments in which relevant differences can instead occur.

We consider in particular the MME in the secular regime, noting that this regime is typically encountered in several contexts such as in quantum optics setups~\cite{breuer}. The steady state $\rho^{\rm sec}$, which satisfies both $\prtq{\rho^{\rm sec},H_S+H_{LS}}=0$  and ${\cal D}^{\rm sec}\prt{\rho^{\rm sec}}=0$, is
\begin{equation}\label{rss}
\rho^{\rm sec}=\frac{\gamma^{\theta}_-}{\gamma^{\theta}_+ +\gamma^{\theta}_-}|\phi_+\rangle \langle \phi_+| +\frac{\gamma^{\theta}_+}{\gamma^{\theta}_++\gamma^{\theta}_-}|\phi_-\rangle \langle \phi_-|,
\end{equation}
where the apex ${\rm ``sec"}$ refers to the  secular master equation.
The collection of steady states that are obtained as a function of the control parameter  $\theta$ and of the  phase $\varphi$ (once it is restored) describes a surface in the Bloch vector representation which is invariant under a rotation around the $z$-axis.

We consider structured environments characterized by a spectral density varying notably  around $\omega_L$ on the scale of the dressed  frequency $\nu$.
In this scenario, even in the limit where the secular approximation holds, the  microscopic approach provides a family of target steady states that may be not close to the ones obtained with the FDME. While in the FD case, there is only one value of the spectral density that matters, the two additional sidebands $\omega_L\pm \nu$ must be considered according to the microscopic derivation (see Eq.~\eqref{gammas}).

When  $n_+ \approx   n_- \approx n_0  \approx n_{\rm fd}$, Eq.~\eqref{rss} takes the form (after restoring the phase $\varphi$)
\begin{eqnarray}\label{ree2}
\rho_{ee}^{\rm sec}&\approx& \frac{n_{\rm fd}}{1+2 n_{\rm fd}}+    \frac{S^2C^2}{1+2 n_{\rm fd}}\frac{S^2 \gamma_-+C^2 \gamma_+}{S^4 \gamma_-+C^4 \gamma_+}, \nonumber \\
\rho_{eg}^{\rm sec}& \approx& \frac{S C}{1+2 n_{\rm fd}}\frac{S^4 \gamma_--C^4 \gamma_+}{S^4 \gamma_-+C^4 \gamma_+} e^{-i \varphi}.
\end{eqnarray}
Exploiting the dependence on the two frequencies $\omega_L \pm \nu$ opens the possibility for taking profit from reservoir engineering. It indeed allows one to deform the ellipsoid of Fig.~\ref{Figsp}, thus  modifying the family of target states. For instance, one of the possible consequences is that the equator of the ellipsoid can be broadened, allowing one to get higher values for the coherence, as it is always possible to reduce the weight of the smaller term in Eq.~\eqref{rss} and then to obtain purer states.

\subsection{The case of zero temperature}\label{SecT0}

 \begin{figure}[t!]
  \centering
  \includegraphics[width=0.49 \textwidth]{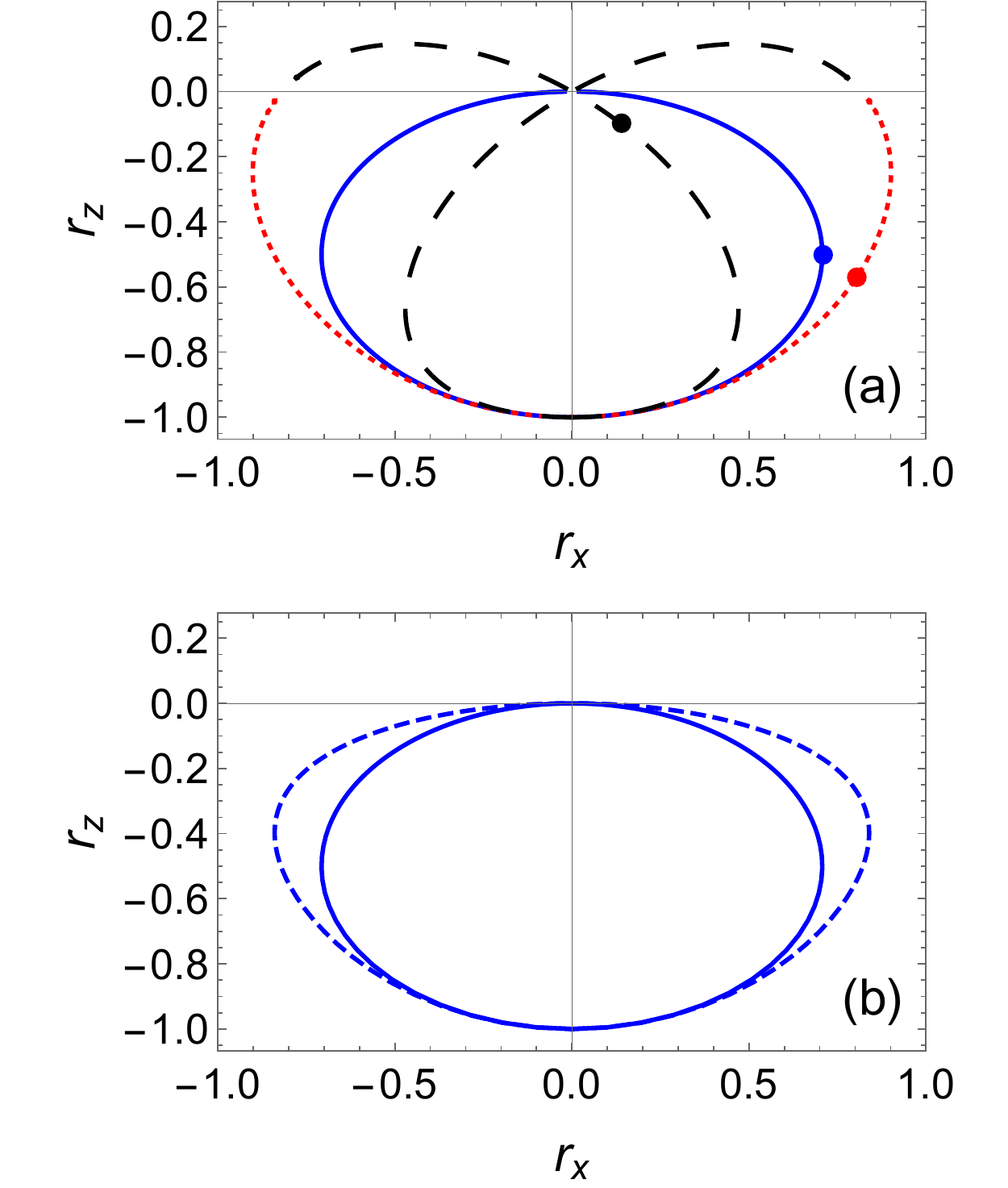}\\
  \caption{Families of steady states (components $x$ and $z$ of the Bloch vector, $r_x$ and $r_z$) determined from the FDME and from the secular MME by varying the control field parameters $\Omega$, assuming values $\ge 0$, and $\varphi$,  being equal to $0$ or $\pi$, for a positive fixed value of $\Delta$. Panel (a): the FDME case  is represented by the blue solid line (the dependence on $\gamma_{\rm fd}$ is assumed to be negligible) while the red dotted and the black dashed lines represent the microscopic steady states when  $x=\gamma_-/\gamma_+$ is kept fixed for any $\nu$ and equal to, respectively, 0.1 and 10. The three enlightened points represent the three steady states obtained by using $\Omega=\sqrt{2}\Delta$ and $\varphi=\pi$, which, in the FDME case, gives the maximum allowed coherence.
   Panel (b): the family of stationary states has been calculated either assuming the FDME (solid line) or the Lorentzian density of states given in Eq.~\eqref{lor} with $\omega_c=\omega_0$ and $\lambda=\Delta $  (dashed line).
  }\label{figeng}
\end{figure}

We start the analysis with the zero-temperature case.
The scenarios where $J(\omega_L+\nu)\gtrless J(\omega_L-\nu)$ are compared with the flat spectral density case in Fig.~\ref{figeng}, where the components $x$ and $z$ of the Bloch vector of the steady states, $r_x=2\, \mathrm{Re}[\rho_{eg}]$ and $r_z=2\rho_{ee}-1$, are plotted.  In Fig.~\ref{figeng}(a), we consider fixed values for the ratio $x=\gamma_-/\gamma_+=J(\omega_L-\nu)/J(\omega_L+\nu)$ for all values of the the dressed energy.  This analysis permits to visualize for any value of the control parameter $\theta$ of Eq.~\eqref{theta} (or equivalently of $\Omega/\Delta$) how much the steady states are expected to differ in the two approaches for a given $x$. In the two panels, all the parts of the different lines are obtained by considering a fixed positive $\Delta$ and using $\Omega$, assuming values $\ge 0$, and $\varphi$, being equal to $0$ or $\pi$, as control parameters. In particular, with respect to the FDME, purer states can be obtained (the coherence may become closer to the maximum allowed value of $1/2$) and even the population inversion can be reached. We observe that the FDME is considered in the case when its dependence on $\gamma_{\rm fd}$ is negligible, and so coincides with the microscopic secular solution in the limit of flat spectrum ($x=1$).

As an example, let us consider the case where the target state reached using the FD dynamics at zero temperature is one of the maximally allowed coherent states $\rho_{\rm mc}^{\rm fd}$ of the ellipsoid in the $z$ basis, that is,  a point that lies on the its equator \cite{sauer,sauer2}. This class of states  is obtained using  $\Omega=\pm\sqrt{2} \Delta$ and, written in the Bloch form, is $\vec{r}_{\rm mc}= \{\mp \cos\varphi/\sqrt{2},\;  \mp \sin  \varphi/\sqrt{2},\;   -1/2    \}$. We focus on the case $\Omega=\sqrt{2} \Delta$ and $ \varphi=\pi$, obtaining then $\vec{r}_{\rm mc}= \{1/\sqrt{2},\;  0,\;   -1/2  \}$.  On the other hand, in the presence of structured reservoir, taking  $\Omega=\sqrt{2} \Delta$ and $ \varphi=\pi$, we would end up in  $\vec{r}\simeq\{ 0.805, 0, -0.569     \}$ using $x=0.1$ or in  $\vec{r}\simeq\{ 0.134, 0 , -0.095  \}$ using $x=10$. The three states, reached with the same control field, are visualized with points in Fig.~\ref{figeng}(a). The distances between
these points clearly point out  how much could be the error of using the FDME to predict the steady state in a given control protocol.

In order to treat a specific physical scenario where $x$ varies when the control field parameters are changed, we consider the case in which the spectral density has the Lorentzian profile
 \begin{equation}\label{lor}
 J_{\rm Lor} (\omega)=\frac{\gamma_l }{2\pi}\frac{\lambda^2}{(\omega_c-\omega)^2+\lambda^2},
 \end{equation}
 where the parameter $\lambda$ defines the width of the curve and $\omega_c$ its center. We note that to satisfy the Markovian approximation used for the derivation of the MME, $\lambda$ must be much greater than $\gamma_l$. The flat spectral density case is recovered in the limit $\lambda\to\infty$. In this case, one can expect that only in some parts of the parameter space the deformation is relevant. On the tails of the curve, we fall for instance in something similar to the flat spectral density case, which gives the same results of the FDME. The differences in the case of a Lorentzian spectral density are depicted in Fig.~\ref{figeng}(b), where we have assumed that the dependence on $\gamma_{\rm fd}$ of the FDME steady solutions is negligible and calculated the steady states by varying $\Omega/\Delta$ in the case of a fixed Lorentzian, with $\lambda=\Delta$ (we have fixed a positive value for $\Delta$ and varied $\Omega$, assuming values $\ge 0$, and $\varphi$, being equal to $0$ or $\pi$).

We have numerically compared the secular MME curve in Fig.~\ref{figeng}(b) with the one obtained adding the non-secular terms at zero temperature and for values of $\gamma_{\rm fd}$ much smaller than $\Delta$. In general, the non-secular curve is very close to the MME curve except when one approaches the origin of the axis, for values of $\Omega$ much larger than $\Delta$. For instance, for $\gamma_{\rm fd}/\Delta = 0.001$  we observe differences for $\Omega/\Delta$ greater than 100. When this ratio overcomes a given value, we observe steady values of $r_x$ different from zero, $r_z$ becoming positive but very close to zero, and at a certain point the non-secular MME starts to predict non-physical steady states. The occurrence of differences between secular and non-secular master equations has been discussed in Ref.~\cite{Shen}.

\begin{figure}[t!]
  \centering
  \includegraphics[width=0.45 \textwidth]{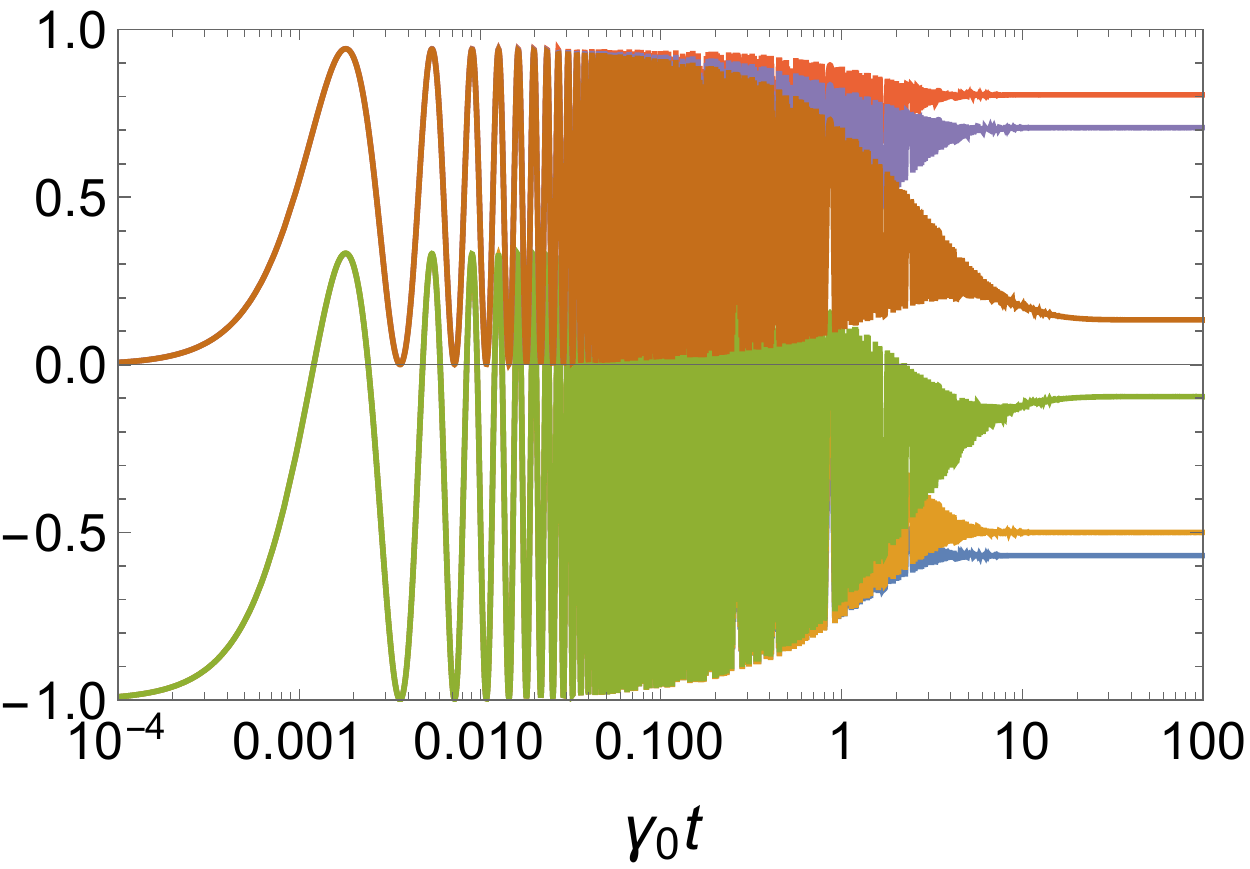}\\
  \caption{The Bloch vector components $r_x$ and $r_z$,  as a function of time (in units of $\gamma_0^{-1}$ and in logarithmic scale). The initial state is $\ket{g}$. Three cases are considered: $x=0.1$ ($\gamma_-=0.2 \gamma_0 $ and $\gamma_+=2 \gamma_0 $), $x=1$ (fixed dissipator, $\gamma_-=\gamma_+= \gamma_0 $) and $x=10$ ($\gamma_-=2 \gamma_0 $ and $\gamma_+=0.2 \gamma_0 $). The dependence of FDME on $\gamma_{\rm fd}$ is assumed to be negligible ($\gamma_{\rm fd}\approx\gamma_{0}=\Delta/1000$). All the $r_x$ curves start from 0 and reach the larger value for the case $x=0.1$, the intermediate one for the case $x=1$ and the lower one for the case $x=10$. All the $r_z$ curves start from -1 and reach the larger value (in modulus) for the case $x=0.1$, the intermediate one (in modulus) for the case $x=1$ and the lower one (in modulus) for the case $x=10$.}\label{dynamics}
\end{figure}

In order to show how different steady states for the same values of the control field parameters are dynamically obtained, we report in Fig.~\ref{dynamics} the time evolution of $r_x$ and $r_z$ for the same cases of the points highlighted in Fig.~\ref{figeng}(a), obtained with $\Omega=\sqrt{2}\Delta$ and $\varphi=\pi$. In particular, we choose $\gamma_-=0.2 \gamma_0 $ and $\gamma_+=2 \gamma_0 $ for the case $x=0.1$, $\gamma_-=\gamma_+= \gamma_0 $ for the fixed dissipator case ($x=1$), and $\gamma_-=2 \gamma_0 $ and $\gamma_+=0.2 \gamma_0 $ for the case $x=10$. As before, the FDME is considered in the case when its dependence on $\gamma_{\rm fd}$ is negligible (we assume $\gamma_{\rm fd}\approx \gamma_{0}=\Delta/1000 $). The qubit is initially in the ground state $\ket{g}$. The values of the rates $\gamma_p$ are chosen without referring to a specific spectral density.

\begin{figure}[t!]
  \centering
  \includegraphics[width=0.45 \textwidth]{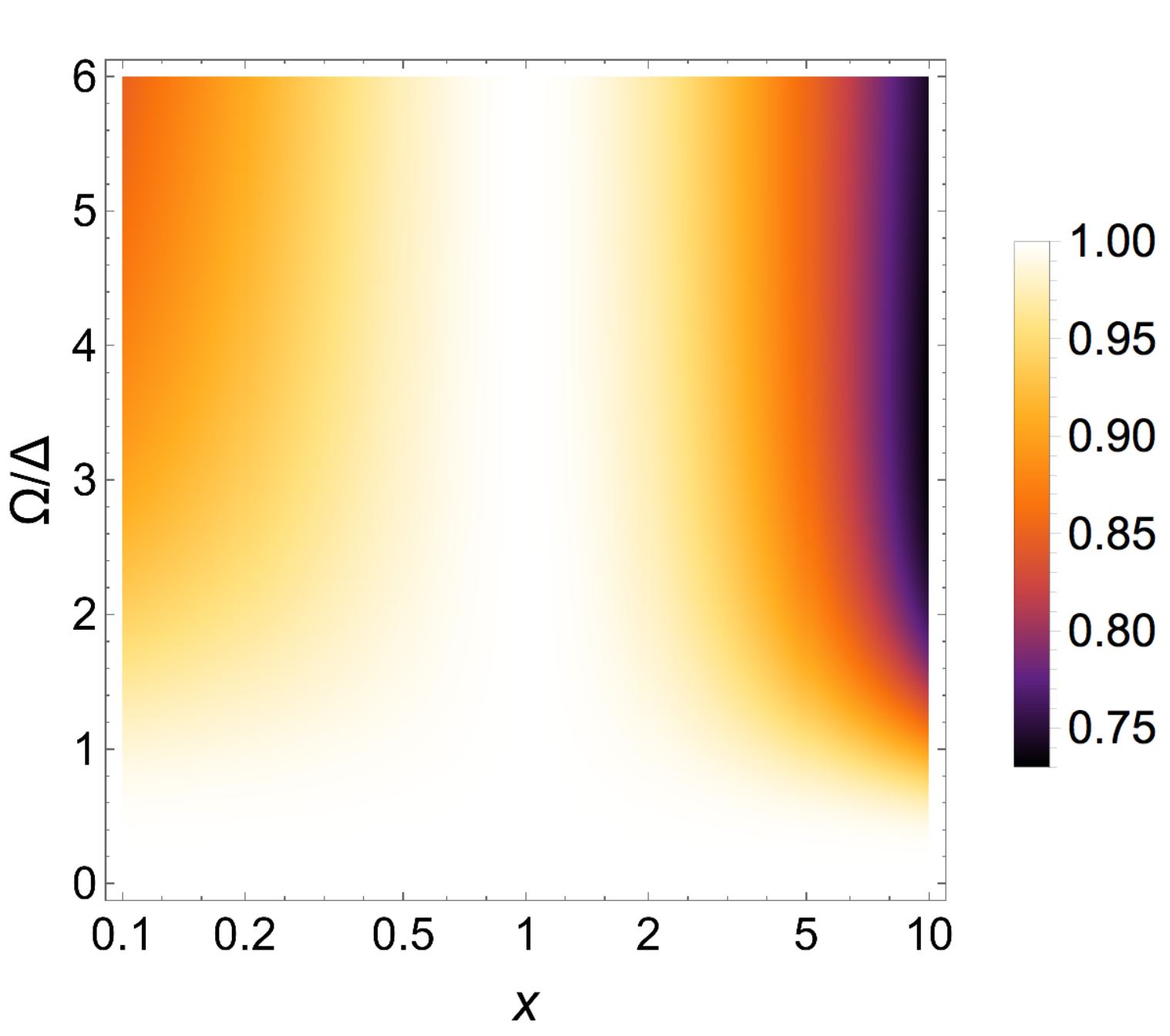}\\
  \caption{Fidelity between the FDME  steady states (their dependence  on $\gamma_{\rm fd}$ is assumed to be negligible) and the ones obtained with the secular MME by releasing the flat-spectrum assumption, as a function of the control field $\Omega/\Delta$ and of the ratio $x=\gamma_-/\gamma_+$ (in logarithmic scale). }\label{engfid}
\end{figure}

We have then shown that, in general, using the FDME can cause a lack of accuracy in determining the steady state, which would be detrimental  in a quantum control protocol.
This effect can be enlightened by considering the distance between the stationary state induced by a structured  spectral density, as predicted by the MME,  and the one given by the FDME as a function of the control field parameter $\Omega/\Delta$. In Fig.~\ref{engfid}, we use the fidelity as a measure of such distance. For two arbitrary states $\rho$ and $\sigma$ it is defined as $\mathrm{Tr} \prtg{\sqrt{\sqrt{\rho} \sigma \sqrt{\rho}}}^2 $. It is important to stress that a fidelity of the order of $3/4$ is already an indication of a dramatic difference between two states. For instance, the fidelity between a two-qubit Bell state and the state obtained from it by removing the coherence is $1/\sqrt{2}$. In Fig.~\ref{engfid}, an important discrepancy may be observed for $\Omega/\Delta \gtrsim 1$. In particular, for a given value of $\Omega/\Delta$, smaller values of fidelity are obtained when $x$ moves away from 1.  The behavior for $\Omega/\Delta < 1$ is instead reminiscent of the fact that for small angles $\theta$ the microscopic dissipator tends to the FD one, as shown before Eq.~\eqref{FD}.

One may raise doubts about the freedom in the choice of the spectral density. In particular, the Markovian approximation could break down for some frequency region. Here, we want to remark that the results of this section hold for  values of the system-bath coupling well behind the weak-coupling limit, such that  the Markovian character of the dynamics is warranted. In any case, even for an intermediate coupling constant, we are interested in the stationary regime, that takes place long after all the possible non-Markovian effects have been washed out.

\subsection{The case of non zero temperature}

According to what said so far, a structured spectral density allows for a broader family of target states but, at the same time, would typically give solutions that are distinct from the ellipsoid predicted by the FDME.
We now show that zero-temperature FDME steady states  can be recovered in the case of a structured environment by exploiting tailored thermal effects.

\begin{figure}%[t!]
  \centering
  \includegraphics[width=0.45 \textwidth]{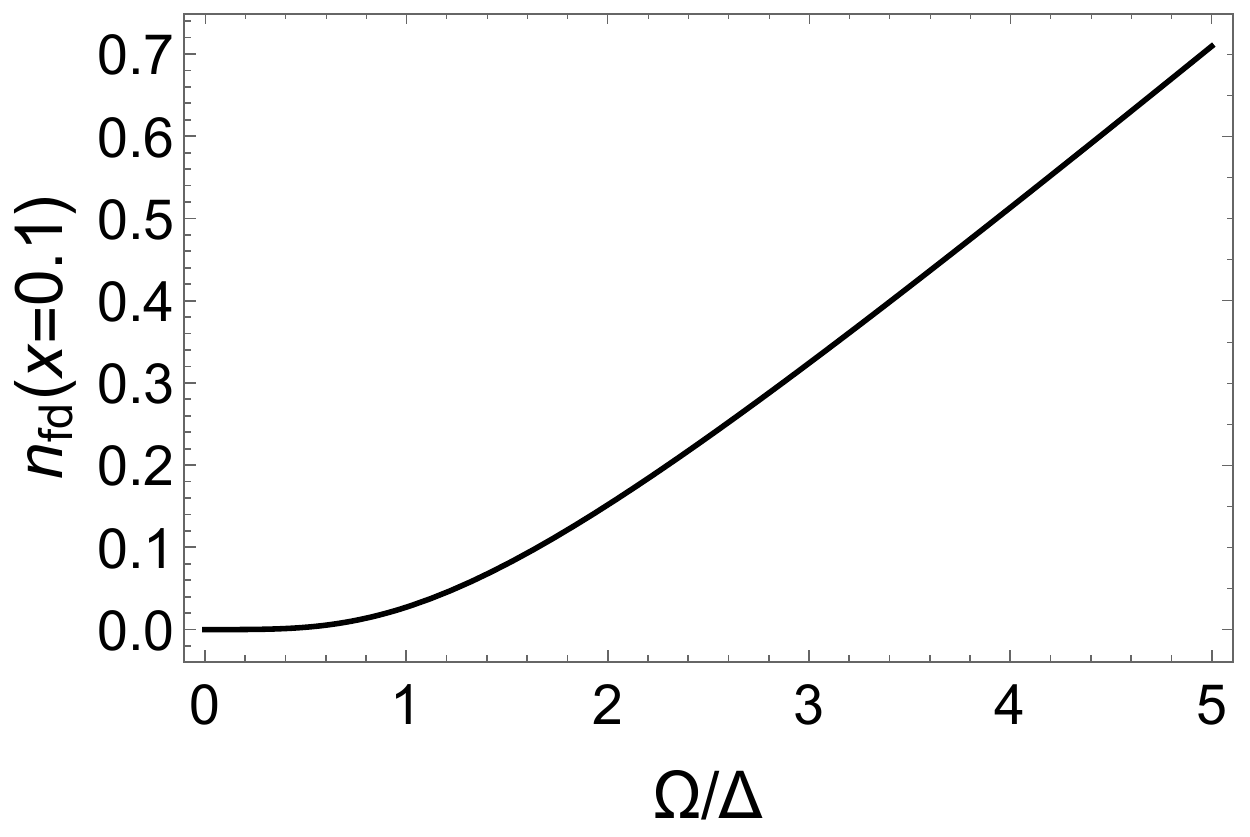}\\
  \caption{Thermal factor $n_{\rm fd}(x=0.1)$ as a function of $\Omega/\Delta$ necessary to compensate the effects due to a ratio $x=\gamma_-/\gamma_+=0.1$ in order for the MME steady state at the temperature corresponding to $n_{\rm fd}(x=0.1)$ to end up in the zero-temperature flat-spectrum ellipsoid formed by the (approximated) FDME steady states.}\label{FigT}
\end{figure}

To this aim, we consider the FDME steady states  of Eq.~\eqref{ssfd} in the limit when the terms depending on $\gamma_{\rm fd}$ are negligible (being always  $n_+ \approx   n_- \approx n_0  \approx n_{\rm fd}$):
\begin{eqnarray}\label{ree}
\rho_{ee}^{\rm sec} &\approx& \frac{n_{\rm fd}}{1+2 n_{\rm fd}}+   \frac{\Omega^2/(1+2 n_{\rm fd})}{4 \Delta^2+2 \Omega^2} , \nonumber \\
\rho_{eg}^{\rm sec}&\approx & -\frac{ \Omega \Delta/(1+2 n_{\rm fd} )  }{2 \Delta^2+ \Omega^2} e^{-i \varphi}.
\end{eqnarray}
We indicate them with apex ${\rm ``sec"}$  since they coincide the with the steady states of the secular MME (see Eq.~\eqref{ree2}) in the limit of flat spectrum ($x=1$). We
compare them at zero temperature with the general case  of Eq.~\eqref{ree2}  that depends both on $n_{\rm fd}$ and on the ratio $x=\gamma_-/\gamma_+$, and look, for any given $x$, for the existence of solutions of
\begin{equation}
\begin{cases}
\rho_{ee}^{\rm sec}(x=1,n_{\rm fd}=0)=\rho_{ee}^{\rm sec}( x,n_{\rm fd})\\
\rho_{eg}^{\rm sec}(x=1,n_{\rm fd}=0)=\rho_{eg}^{\rm sec}(x,n_{\rm fd})
 \end{cases}.
\end{equation}
The solution of both equations is given by
\begin{equation}
n_{\rm fd}(x)=\frac{C^4 S^4(1-x)}{(C^2 -S^2)(C^4+x S^4)}.
\end{equation}
Solutions corresponding to physical values of $n_{\rm fd}$ (that is $n_{\rm fd}\ge 0$) only appear for $0\le x\le 1$, which is easy to understand looking at Fig.~\ref{figeng}(a). In fact, thermal effects are expected to reduce the coherence of any state, making it impossible to move from the black line ($x=10$) to the blue one ($x=1$). The behavior of $n_{\rm fd}(x=0.1)$ is plotted in Fig.~\ref{FigT} as a function of the control field parameter $\Omega/\Delta$.  The needed thermal correction is very small as long as $\Omega \lesssim \Delta$, as the same argument used to explain the behavior observed in Fig.~\ref{engfid} holds.

\section{Discussion and conclusions}\label{sec5}
Master equations are a powerful tool to analyze the dissipative dynamics of quantum systems. They are usually obtained by making a series of assumptions that need to be fulfilled and to be verified in realistic setups, as, in general, exact solutions are not available.  They are often introduced based on  phenomenological assumptions. Here, we have derived a microscopic master equation for a driven qubit and compared it with the fixed dissipator model, which is widely used especially in the quantum control community, as it allows to explore the behavior of entire families of control Hamiltonians in a simple way.
 We have found that, in the weak-coupling regime, the steady states of the two approaches can be very different in the case of a structured environment, while they are practically identical for a flat spectrum.

In conclusion,  considering the simplest case of a driven qubit, we have assessed the limit of validity of the phenomenological approach for the specific task on asymptotic time scales. We have explored the possibility of implementing reservoir engineering techniques to widen the family of target states, which are correctly predicted by using microscopic master equations.

Quantum control protocols most often use time dependent fields, implying time dependent Rabi frequency, detuning, and phase as control parameters. For slowly varying parameters, one expects that the FDME and MME still coincide for a flat environment spectrum, and that the difference between them still persists for structured environments. The expected rich variety of target states  resulting from structured environments could then be exploited using microscopic models in quantum control and reservoir engineering schemes.

\begin{acknowledgments}
This work was mainly supported by the French ``Investissements d'Avenir'' program, project ISITE-BFC (contract ANR-15-IDEX-03). G.L.G. acknowledges financial support from the CAIB postdoctoral program. A.S. and S.G. acknowledge additional support from the European Union's Horizon 2020 research and innovation program under the Marie Sklodowska-Curie grant agreement No. 765075 (LIMQUET). G.L.G. and B.B. thank Hans-Rudolf Jauslin, Axel Kuhn, and David Viennot for useful discussions. B.B. thanks Mauro Paternostro for helpful comments..
\end{acknowledgments}

\appendix*
    \section{Master equation}\label{appe}
In this appendix, we derive the microscopic master equation of the driven qubit, presenting its various parts.
Using Eq.~\eqref{sx}, in the interaction picture with respect to $H_S + H_E$, the interaction Hamiltonian of Eq.~\eqref{hi} reads
\begin{equation}\label{me}
\tilde  H_I(t)=B(t)\left(f_+^t\tilde{\sigma}_++f_-^t\tilde{\sigma}_-+f_z^t\tilde{\sigma}_z\right),
\end{equation}
with $B(t)=\sum_k  \hbar\prt{g_k a_k e^{-i \omega_k t}+  g_k^* a_k^\dag e^{i \omega_k t}}$ and where
\begin{eqnarray}\label{fs}
f_+^t&=&e^{i \nu t}\prtq{C^2 e^{i(\omega_L t+\varphi)}-S^2 e^{-i(\omega_L t+\varphi)}},\nonumber\\
f_-^t&=&e^{-i \nu t}\prtq{C^2 e^{-i(\omega_L t+\varphi)}-S^2  e^{i(\omega_L t+\varphi)}},\nonumber\\
f_z^t&=& S C\prtq{e^{i(\omega_L t+\varphi)}+e^{-i(\omega_L t+\varphi)}}.
\end{eqnarray}
The operators entering the master equation are multiplied by $f_i^t f_j^{t-s} $, with $i,j=+,-,z$. Thus,  there will be secular terms for $\{i, j\}$ such that $f_i^t= f_j^{t*} $ and non-secular terms in all other cases.
In general, the products $f_i^t f_j^{t-s} $ may have parts oscillating at the laser frequency  $e^{\pm 2 i \omega_L t}$.
For instance,
\begin{eqnarray}\label{fp}
 f_+^t f_+^{t-s}&=& {\Big \{ }  C^4 e^{i[\omega_L(2t-s)+2\varphi]}+S^4 e^{-i[\omega_L(2t-s)+2\varphi]} \nonumber     \\   &-&   2  C^2 S^2  \cos\omega_L s  {\Big \}}  e^{i\nu(2 t-s)}.
\end{eqnarray}
On the basis of the condition assumed in Sec.~\ref{sec2}, $\omega_L\gg \Delta,\Omega$, we note that the first two fast-oscillating terms in Eq.~\eqref{fp} can be neglected. Neglecting this kind of terms  is completely equivalent to obtain the master equation writing the  interaction Hamiltonian in rotating wave approximation:
\begin{equation}\label{hi2}
H_I=\sum_k \hbar \prtq{ g_k a_k e^{i (\omega_L t+\varphi)}\sigma_+ + g_k^* a_k^\dag e^{-i (\omega_L t+\varphi)}\sigma_-}.
\end{equation}
In this limit, the products linked to the secular terms
\begin{eqnarray}\label{fs2}
 f_-^t f_+^{t-s}&\approx & e^{-i\nu s} \prtq{C^4 e^{-i \omega_L s}+ S^4  e^{i \omega_L s}}, \nonumber \\
 f_+^t  f_-^{t-s}&\approx & e^{i\nu s} \prtq{ C^4 e^{i \omega_L s}+ S^4  e^{-i \omega_L s}},\nonumber \\
  f_z^t f_z^{t-s}&\approx &   2 S^2 C^2\cos \omega_L s,
\end{eqnarray}
determine the coefficients of Eq.~\eqref{gammas}. Non-secular terms are determined by the products
\begin{eqnarray}
 f^t_+f^{t-s}_+&\approx&-2 C^2 S^2 e^{i\nu(2t- s)} \cos\omega_L s, \nonumber\\
    f^t_+f^{t-s}_z&\approx&C S\prt{C^2  e^{i\omega_L s}-S^2       e^{-i\omega_L s}}e^{i \nu t},  \nonumber  \\
    f^t_zf^{t-s}_+&\approx&  e^{i\nu(t- s)}  CS \prt{C^2  e^{-i\omega_L s}-S^2  e^{i\omega_L s}},
\end{eqnarray}
together with $ f^t_-f^{t-s}_-=\prt{ f^t_+f^{t-s}_+}^*$,  $f^t_-f^{t-s}_z=\prt{f^t_+f^{t-s}_z}^*$, and $f^t_zf^{t-s}_-=\prt{f^t_z f^{t-s}_+}^*$. The factors  $e^{\pm i \nu t}$ and $e^{\pm 2 i \nu t}$ disappear once one moves back to the Schr\"odinger picture.  We indicate with $\overline{f^t_i f^{t-s}_j}$ the products $f^t_i f^{t-s}_j$  after discarding   the factors $e^{\pm i \nu t}$ and $e^{\pm 2 i \nu t}$ and, taking the continuum limit, we introduce the spectral density  $J(\omega)=\sum_k |g_k|^2\delta(\omega-\omega_k)$, such that the trace over the bath's degrees of freedom is transformed into an integral over all the frequencies.
The Born-Markov master equation, assuming a factorized initial condition for the system and its bath, is then given by  \cite{breuer, gardiner}
\begin{eqnarray}\label{ME2}
 &&  \!\!\!\!\! \dot \rho=-\frac{i}{\hbar}[H_S, \rho] +\frac{1}{\hbar^2}\sum_{i,j=+,-,z} \int_0^{\infty} ds   \nonumber\\  &&  \!\!\!\!\!  \left[\overline{f^{t*}_i f^{t-s}_j}
 \langle B(t)B(t-s)\rangle  \prt{\tilde{\sigma}_j \rho\tilde{\sigma}_i^\dag - \tilde{\sigma}_i^\dag\tilde{\sigma}_j \rho}+\rm h.c. \right],\qquad
\end{eqnarray}
where h.c. denotes Hermitian conjugation and the bath correlation functions, taking a thermal equilibrium state $\rho_B$ at temperature $T$, are given by
\begin{eqnarray} \label{funccorr}
&&{\rm Tr}_B\{ B(t)B(t-s)\rho_B  \}=\hbar^2 \int_0^{\infty} d\omega J(\omega)\nonumber \\  &&\quad \times \prtg{[1+ n(\omega)] e^{-i \omega s }
 +n(\omega) e^{i \omega s }}.
\end{eqnarray}
The explicit development of Eq.~\eqref{ME2} leads to Eq.~\eqref{metot}. In particular, in order to calculate the coefficients of the master equation one makes use of the identity
\begin{equation}\label{expintegral}
\int_0^\infty e^{\pm i \varepsilon s}d s=\pi \delta(\varepsilon)\pm i \mathcal{P}\frac{1}{\varepsilon},
\end{equation}
where  $\delta(\varepsilon)$ is the Dirac delta function and $\mathcal{P}$ denotes the Cauchy principal value.

The Lamb shift Hamiltonian of Eq.~\eqref{metot} is  given by
\begin{equation}\label{Lamb-shift Hamiltonian}
H_{LS}= \hbar \prt{s_+  \tilde{\sigma}_+\tilde{\sigma}_-+s_-  \tilde{\sigma}_-\tilde{\sigma}_++ s_z  \tilde{\sigma}_z^2},
\end{equation}
where
\begin{eqnarray}
 s_+ &=& \mathcal{P} \int_0^{\infty}   d\omega  J (\omega) \left\{ \frac{C^4 \prtq{1+n(\omega)}}{(\omega_L+\nu)-\omega}  -  \frac{S^4 n(\omega)}{(\omega_L-\nu)-\omega} \right\},\nonumber \\
 s_-  &=& \mathcal{P} \int_0^{\infty}   d\omega  J (\omega) \left\{\frac{S^4 \prtq{1+n(\omega)}}{ (\omega_L-\nu)-\omega}-\frac{C^4 n(\omega)}{(\omega_L+\nu)-\omega}\right\},\nonumber \\
 s_z &=&\mathcal{P} \int_0^{\infty}   d\omega  J (\omega) S^2C^2\frac{1}{\omega_L- \omega} .
\end{eqnarray}
In the secular limit, it holds $[H_S,H_{LS}]=0$.

As for the non-secular part ${\cal D}^{\rm nsec}(\rho)$ we have
\begin{eqnarray}\label{Non secular part}
{\cal D}^{\rm nsec}(\rho)&=&\prt{\gamma_{++}^\theta{+is_{++}}}\tilde{\sigma}_+\rho \tilde{\sigma}_+ \nonumber\\
&+& \prt{\gamma_{+z}^\theta{+is_{+z}} }\prt{ \tilde{\sigma}_+ \tilde{\sigma}_z \rho-\tilde{\sigma}_z \rho\tilde{\sigma}_+}\nonumber\\
&+& \prt{\gamma_{-z}^\theta{+is_{-z}}} \prt{ \tilde{\sigma}_- \tilde{\sigma}_z \rho-\tilde{\sigma}_z \rho\tilde{\sigma}_-}\nonumber\\
&+& \prt{\gamma_{z+}^\theta {+is_{z+}}}\prt{ \tilde{\sigma}_z \tilde{\sigma}_+ \rho-\tilde{\sigma}_+ \rho\tilde{\sigma}_z}\nonumber\\
&+& \prt{\gamma_{z-}^\theta{+is_{z-}}} \prt{ \tilde{\sigma}_z \tilde{\sigma}_- \rho-\tilde{\sigma}_- \rho\tilde{\sigma}_z}\nonumber\\
&+&{\rm h.c.},
\end{eqnarray}
where the various coefficients  $\gamma_{ij}^\theta$ and $s_{ij}$ can be computed by explicitly developing Eq.~\eqref{ME2}:

\begin{eqnarray}\label{nonsgammas}
\gamma_{++}^\theta&=& -\frac{1}{2} C^2 S^2 \prtq{\gamma_-(1+2n_-) + \gamma_+(1+2n_+)}, \nonumber\\
\gamma_{z+}^\theta&=&  -\frac{1}{2} C S\left[\gamma_+ n_+ C^2 -\gamma_-(1+n_-)S^2  \right],\nonumber\\
\gamma_{z-}^\theta&=& -\frac{1}{2} C S\left[  \gamma_+(1+n_+)C^2 -\gamma_-n_-S^2  \right],\nonumber\\
\gamma_{+z}^\theta&=&-\frac{1}{2}  \gamma_0C S\left[(1+n_0)C^2 -n_0 S^2 \right],\nonumber\\
\gamma_{-z}^\theta&=&-\frac{1}{2}  \gamma_0 C S\left[n_0C^2 -(1+n_0) S^2 \right],
\end{eqnarray}
and
\begin{eqnarray}\label{nsprincipal parts}
s_{++}&=&  -\mathcal{P}\int_0^{\infty}   d\omega  J (\omega) C^2 S^2\nonumber\\
&&\times\left[ \frac{1+2n(\omega)}{(\omega_L-\nu)-\omega}  -  \frac{ 1+2n(\omega)}{(\omega_L+\nu)-\omega} \right], \nonumber \\
s_{z+}&=& \mathcal{P}\int_0^{\infty}   d\omega  J (\omega) C S\nonumber\\
&&\times\left\{ \frac{S^2[1+n(\omega)]}{(\omega_L-\nu)-\omega}  +  \frac{C^2n(\omega)}{(\omega_L+\nu)-\omega} \right\},\nonumber\\
s_{z-}&=& -\mathcal{P}\int_0^{\infty}   d\omega  J (\omega) C S\nonumber\\
&&\times\left\{ \frac{S^2n(\omega)}{(\omega_L-\nu)-\omega}  +  \frac{C^2[1+n(\omega)]}{(\omega_L+\nu)-\omega} \right\},\nonumber\\
s_{+z}&=&-\mathcal{P}\int_0^{\infty}   d\omega  J (\omega) C S\nonumber\\
&&\times\left\{ \frac{C^2[1+n(\omega)]}{\omega_L-\omega}  +  \frac{S^2n(\omega)}{\omega_L-\omega} \right\},\nonumber\\
s_{-z}&=&\mathcal{P}\int_0^{\infty}   d\omega  J (\omega) C S\nonumber\\
&& \times \left\{ \frac{C^2n(\omega)}{\omega_L-\omega}  +  \frac{S^2[1+n(\omega)]}{\omega_L-\omega} \right\}.
\end{eqnarray}
For each pair of $i,j$ in Eq.~\eqref{ME2} the part of the integrals involving the delta function, gives us the decay rates of Eq.~\eqref{gammas} when $i=j$ and the ones of  Eq.~\eqref{nonsgammas} when $i\neq j$, for any spectral density. The principal part in Eq.~\eqref{expintegral} leads to the Lamb shift Hamiltonian of Eq.~\eqref{Lamb-shift Hamiltonian} and  the terms in Eq.~\eqref{nsprincipal parts}. It can be shown (see for instance Ref.~\cite{tanas}) that, in the case of a flat spectral density, all of these principal parts vanish.
This can be obtained by firstly performing the integrals by using a Lorentzian spectral density, and by then taking the width of this Lorentzian to infinity. In the case of a non-flat spectrum, we treat these terms taking again the Lorentzian spectral density. Since in the secular MME these terms lead to the Lamb shift Hamiltonian, which is nothing but energy shift, their effect is not relevant for the steady states, while in the case of the non-secular MME their contribution, in general, can not be neglected.

Finally, keeping the terms~\eqref{Non secular part} in Eq.~\eqref{metot}, it is possible to show that in the flat-spectrum limit, under the approximation $n_+ \approx   n_- \approx n_0  \approx n_{\rm fd}$, the non-secular MME gives exactly the same result as FDME, i.e, using $H=H_S$, Eq.~\eqref{metot} becomes Eq.~\eqref{MEFD} for any $\gamma_{\rm fd}$.

%\onecolumngrid
%\vfill\clearpage
%\twocolumngrid
%\setcounter{equation}{ 0}

\end{document}